# AN INTERNAL WAVE AS A FREQUENCY FILTER FOR SURFACE GRAVITY WAVES ON WATER

Lossow Konstantin , PhD, Lossow Nikolay, PhD.
The Department of Higher Mathematics at Moscow State University of Geodesy and Cartography.

**Introduction.** In this article we consider a one-dimensional model of the interaction between surface and the internal gravity waves on water. The internal wave is modeled by a flow. Using the conventional method of stationary phase, it is found approximation for the height of the surface wave on the flow by the " elementary quasi stationary " solutions (each such solution corresponds to its own frequency). We show that the flow acts as a frequency filter for gravity waves. Many researchers considered this problem of the interactions: Phillips [1], Longuet-Higgins [2], Smith [3], McKee [4], Hughes [5], … Here we use an approach which was described earlier by Hughes [5]. The surface wave field is assumed to be irrotational, the potential is expressed as a Fourier integral and the equations of motion are linearized. The internal wave is modeled by its basic form: a non-dispersive field with a horizontal current that is uniform over all depth insignificantly affected by the surface waves. We ignore surface tension and wind growth/decay effects. The depth is infinite.

## § 1 One - dimensional model of the interaction of surface gravity waves with the flow.

We use the standard Laplace equation with dynamic and kinematic boundary conditions at a free water surface,

$$\Delta^2 \Phi = 0 ,$$

$$\partial \Phi / \partial t + (1/2) \cdot q^2 + g \cdot z = 0 , \qquad (1)$$

at $z = \zeta$

$$\partial \zeta / \partial t + \partial \Phi / \partial x \cdot \partial \zeta / \partial x - \partial \Phi / \partial z = 0 , \qquad (2)$$

where $\Phi$ is the velocity potential, $\zeta$ is the water height above an arbitrary reference level; $\mathbf{q}$ is the velocity vector $q = |\mathbf{q}|$ ; $g$ is the acceleration due to gravity.

Let the velocity potential and water height be the sums

$$\Phi = \Phi_S + \Phi_I , \qquad \zeta = \zeta_S + \zeta_I$$

where $\Phi_S$ and $\Phi_I$ are the velocity potentials of surface and internal waves; $\zeta_S$ and $\zeta_I$ are the relevant water height, then we have, for the internal wave, the boundary conditions

$$\partial \Phi_I / \partial t + (1/2) \cdot q_I^2 + g \cdot z_I = 0 , \qquad (3a)$$

$$\text{at } z = \zeta_I$$

$$\partial \zeta_I / \partial t + \partial \Phi_I / \partial x \cdot \partial \zeta_I / \partial x - \partial \Phi_I / \partial z = 0 , \qquad (3b)$$

where $\mathbf{q_I}$ is the velocity vector of internal wave, $q_I = |\mathbf{q_I}|$.
Let expand (1) and (2) about $z = \zeta_I$, and ignoring terms of the order $O(\zeta^n{}_S \Phi^m{}_S)$, where $n + m > 1$; assuming the absence of vertical current in the internal wave $W \equiv 0$, and $U \neq 0$ to be the corresponding horizontal current, we find

$$\Delta^2 \Phi_S = 0 ,$$

$$[(\partial/\partial t + U \cdot \partial/\partial x) \cdot \Phi_S]_{\zeta_I} = 0 , \qquad (3c)$$

$$(\partial/\partial t + U \cdot \partial/\partial x)_{\zeta_I} \cdot \zeta_S + (\partial U/\partial x)_{\zeta_I} \cdot \zeta_S + (\partial \zeta_I/\partial x \cdot \partial \Phi/\partial x)_{\zeta_I} = 0 .$$

Subscript $\zeta_I$ indicates that quantities depending on z are to be evaluated at $z = \zeta_I(x, t)$.
  The first of these three equations is solved in an explicit form by an integral transform:

$$\Phi_S = {}_{-\infty}\!\int^\infty {}_{-\infty}\!\int^\infty \Phi(k, \sigma) \cdot \exp[i \cdot k \cdot x - i \cdot \sigma \cdot t + z \cdot k] \cdot dk \, d\sigma ,$$

where k and $\sigma$ are the wave number and frequency of plane surface wave. Denoting the internal wave number by $k_I$ and introducing the ratio of wave numbers as parameter $\mu = k / k_I$, we assume then it to be large $\mu \gg 1$, we retain terms which are $O(1)$ and $O(\mu^{-1})$ only. Some algebraic rearranging results in the expression for wave height,

$$\zeta_S = {}_{-\infty}\!\int^\infty {}_{-\infty}\!\int^\infty (1/g) \cdot [i \cdot \sigma + i \cdot U \cdot k] \cdot \Phi \cdot \exp[i \cdot (k \cdot x - i \cdot \sigma \cdot t + \zeta_I \cdot k] \cdot dk \, d\sigma , \qquad (4a)$$

and the relevant condition of the free surface,

$$_{-\infty}\!\int^\infty {}_{-\infty}\!\int^\infty \Phi \cdot \exp[i \cdot (k \cdot x - i \cdot \sigma \cdot t + \zeta_I \cdot k] \cdot \{g \cdot k - (\sigma + U \cdot k)^2 +$$

$$+ [i \cdot (\sigma + U \cdot k)(\partial U/\partial x + 2 \cdot i \cdot k \cdot U \cdot \partial U/\partial x)] \cdot dk \, d\sigma . \qquad (4b)$$

  Components of the surface wave potential are to be considered as smoothly varying in the $(k, \sigma)$ - space, ie

$$\Phi = a(k, \sigma) \cdot \exp[i \cdot \mu \cdot f(k, \sigma)] + O(\mu^{-1}) , \qquad (5)$$

where phase f and amplitude a are quantities of the order of $O(1)$. Further, we shall consider the special case of one – dimensional flow $U = U(x - c \cdot t)$, induced by internal wave, where $c = \text{const}$ and is the phase speed of internal wave, also let $U \leq 0$ for definiteness. Then we transform the coordinate frame into the moving one,

$x - c \cdot t = \eta$, with $\sigma - k \cdot c = \omega$ as an observable frequency,

$$\partial U/\partial x = dU/d\eta = - (1/c) \cdot \partial U/\partial t .$$

Then (5) becomes

$$\Phi = a(k, \omega) \cdot \exp[i \cdot \mu \cdot f(k, \omega)] + O(\mu^{-1}) ,$$

and from (4a), (4b) we can arrive at

$$\zeta_S = - (i/g) \cdot \int_{-\infty}^{\infty}\int_{-\infty}^{\infty} [k \cdot (U - c) - \omega] \cdot a(k, \omega) \cdot \exp\{i \cdot \mu \cdot [k \cdot x/\mu - \omega \cdot t/\mu + f(k, \omega)]\} \cdot$$

$$\cdot \exp\{\zeta_I \cdot k\} \cdot dk \, d\omega , \qquad (4a')$$

$$\int_{-\infty}^{\infty}\int_{-\infty}^{\infty} a(k, \omega) \cdot \exp\{\zeta_I \cdot k\} \cdot \exp\{i \cdot \mu \cdot [k \cdot x/\mu - \omega \cdot t/\mu + f(k, \omega)]\} \cdot \{g \cdot k - [k \cdot (U - c) - \omega]^2 +$$

$$+ i \cdot [(k \cdot (U - c) - \omega) \cdot \partial U/\partial x + 2 \cdot k \cdot c \cdot \partial U/\partial x + 2 \cdot i \cdot k \cdot U \cdot \partial U/\partial x )]\} \cdot dk \, d\omega . \qquad (4b')$$

According to Hughes result ( [5], formulas (6a), (6b) ) from (4a'), we have

$$g \cdot k = [k \cdot (U - c) - \omega]^2 , \qquad (5a)$$

$$a(k, \omega) = P(\omega) \cdot \{f_{kk}^{//}(k, \omega)/[g - 2 \cdot (-\omega + k \cdot (U - c)) \cdot (U - c)]\}^{1/2} \cdot \exp\{-\zeta_I \cdot k\} , \qquad (5b)$$

where the argument of U is $- \mu \cdot f_k^{/}(k, \omega)$, and $P(\omega)$ is an unspecified "constant" of integration. Therefor substituting (5b) into (4a'), we find

$$\zeta_S = - (i/g) \cdot \int_{-\infty}^{\infty}\int_{-\infty}^{\infty} [k \cdot (U - c) - \omega] \cdot P(\omega) \cdot$$

$$\cdot \{f_{kk}^{//}(k, \omega)/[g - 2 \cdot (-\omega + k \cdot (U - c)) \cdot (U - c)]\}^{1/2} \cdot$$

$$\cdot \exp\{i \cdot \mu \cdot [k \cdot x/\mu - \omega \cdot t/\mu + f(k, \omega)]\} \cdot \exp\{-k \cdot (\zeta_I \cdot (k, \omega) - \zeta_I(x))\} \cdot dk \, d\omega . \qquad (6)$$

If we denote the total phase function as

$$S(k, \omega) = k \cdot x/\mu - \sigma \cdot t/\mu + f(k, \omega) , \qquad (7)$$

stationary points of $S(k, \omega)$ are defined by

$$f_k^{/} = - x/\mu , \qquad (8a)$$

$$f_\omega^{/} = t/\mu . \qquad (8b)$$

In a small segment of x, we can approximate the current function U = U(x) by the dependency

$$U = c \cdot s \cdot (1 + x/x_0), \qquad (9)$$

where $s < 0$, $0 < x_0 > -x$ (see fig. I).

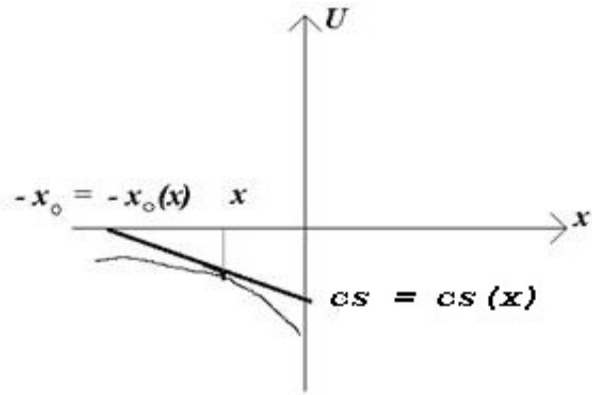

fig. I   the currents flow function graph

The initial phase function f(k, ω) can be in turn determined from (5a) and (8a) via

$$f_k'(k, \omega) = (x_0/\mu) \cdot [1 - 1/s + ((g \cdot k)^{1/2} - \omega)/c \cdot s \cdot k],$$

and, after integration over k, we obtain

$$f(k, \omega) = (x_0/\mu \cdot c \cdot s) \cdot \{(s - 1) \cdot k \cdot c + 2 \cdot (g \cdot k)^{1/2} - \omega \cdot \ln k\} + (x_0/\mu \cdot c \cdot s) \cdot \widetilde{P}(\omega), \quad (10)$$

where $|x_0/c \cdot s| = O(\mu)$, $\widetilde{P}(\omega)$ is an arbitrary function of ω.

## § 2 " Elementary quasi stationary " solution.

Using the standard method of stationary phase in two variables (in our case is k and ω), we find a solution that satisfies, asymptotically, the stationary conditions

$$\omega \to \omega_0,$$
k is bounded,
$$t \to \infty,$$

which we will refer to as the "elementary solution" of system (3c).

The determinant of the form S″ (see, (7), (10)) is

$$\det S'' = (x_0/\mu \cdot c \cdot s \cdot k)^2 \cdot [(\omega - (1/2) \cdot (g \cdot k)^{1/2}) \cdot \widetilde{P(\omega)}_{\omega\,\omega} - 1].$$

Then from (4a') it is easy to find that

$$\zeta_S \approx (-i/g) \cdot k^{3/4} \cdot (x_0/\mu \cdot c \cdot s)^{1/2} \cdot \exp[i \cdot \mu \cdot S(k, \omega) + i \cdot (\pi/4) \cdot \mathrm{sgn} S''] \cdot$$

$$\cdot P(\omega_0, \omega)/|[\omega - (1/2) \cdot (g \cdot k)^{1/2}] \cdot \widetilde{P(\omega)}_{\omega\,\omega} - 1|^{1/2}$$

where $P(\omega_0, \omega)$ is an arbitrary smooth function of two variables, therefore

$$|\zeta_S| \approx (1/g) \cdot k^{3/4} \cdot (-x_0/\mu \cdot c \cdot s)^{1/2} \cdot |P(\omega_0, \omega)|/|[\omega - (1/2) \cdot (g \cdot k)^{1/2}] \cdot \widetilde{P(\omega)}_{\omega\,\omega} - 1|^{1/2}$$

The behavior of $|\zeta_S|$ is governed by the relation

$$A(\omega_0, \omega) = P(\omega_0, \omega)^2/|[\omega - (1/2) \cdot (g \cdot k)^{1/2}] \cdot \widetilde{P(\omega)}_{\omega\,\omega} - 1|.$$

We are to distinguish between two cases.
Case I,

$$\omega \to \omega_0,$$

$$\omega - (1/2) \cdot (g \cdot k)^{1/2} \to 0, \text{ as } t \to \infty.$$

Using (5a), we find

$$\omega_0 = g/[4 \cdot (c - U(x))].$$

The solution squared is determined mainly by the ratio

$$P(\omega_0, \omega)^2/|(\omega - \omega_0)^{1/2} \cdot \widetilde{P(\omega)}_{\omega\,\omega} - 1|,$$

Due to singularity of $P(\omega_0, \omega)$ at $\omega = \omega_0$, we conclude that the squared wave height is in fact determined by

$$P(\omega_0, \omega)^2/|(\omega - \omega_0)^{1/2} \cdot \widetilde{P(\omega)}_{\omega\,\omega}|.$$

For the bounded solution, we should suppose that

$$P(\omega_0, \omega)^2 = (\omega_0 - \omega)^{1/2} \cdot \widetilde{P(\omega)}_{\omega\,\omega} \cdot Q(\omega_0, \omega),$$

where $Q(\omega_0, \omega)$ is a smooth function of two variables.
Case II,

$$\omega \to \omega_0,$$

$$\omega - (1/2)\cdot(g\cdot k)^{1/2} \;-\!/\!\!\to\; 0, \text{ as } t \to \infty.$$

In this case the behavior of wave height $|\zeta(x,t)|$ is determined by the ratio

$$B(\omega_0, \omega) = P(\omega_0, \omega)^2/|[\omega - (1/2)\cdot(g\cdot k)^{1/2}]\cdot \widetilde{P}(\omega)_{\omega\,\omega}|.$$

The choice of $P^2(\omega_0, \omega)$ made in case I leads to a vanishing solution $B(\omega_0, \omega) \to 0$, as $\omega \to \omega_0$.

Another choice of $P^2(\omega_0, \omega)$ leads to $|\zeta(x,t)| \to 0$, as $\omega \to \omega_0$ or singularity existing in case one. So, with no loss of generality we can define

$$\widetilde{P}(\omega)_{\omega\,\omega} = \ln(\exp[\omega_0 - \omega] - 1).$$

From (8b), (10) it follows that

$$\omega = \omega_0 - \ln(1 + k\cdot\exp[c\cdot s\cdot t/x_0]).$$

Then the elementary solution is

$$\zeta_S(x, t, \omega_0) \approx (-\mu\cdot U'(x))^{-1/2}\cdot k^{3/4}\cdot\exp[i\cdot\mu\cdot S(k, \omega)]\cdot$$
$$\cdot P(\omega_0, \omega)/|[\omega - (1/2)\cdot(g\cdot k)^{1/2}]\cdot \widetilde{P}_{\omega\,\omega}(\omega_0, \omega) - 1| + O(\mu^{-1}),$$

where wave parameters $\omega(x,t)$ and $k(x,t)$ are determined by (8a), (8b)

$$S(k, \omega) = k\cdot x/\mu - \omega\cdot t/\mu + f(k, \omega) = k\cdot x/\mu - \omega\cdot t/\mu + (\mu\cdot U'(x))^{-1}\cdot$$
$$\cdot[(s(x) - 1)\cdot k\cdot c - 2\cdot(g\cdot k)^{1/2} - \omega\cdot\ln(k) + \widetilde{P}(\omega_0, \omega)],$$

and $s(x) = -x\cdot U'(x) + U(x)$.

An asymptotic solution for system (3c) is

$$\zeta_S(x, \omega_0) \approx (-\mu\cdot U'(x))^{-1/2}\cdot K_0^{3/4}(x)\cdot Q(\omega_0, \Omega_0(x))\cdot\exp[i\cdot\mu\cdot S(K_0(x), \Omega_0(x))], \text{ as } t \to \infty,$$

where $Q(\omega_0, \omega)$ is a smooth function of two variables which is determined explicitly by initial and boundary conditions and

$$K_0(x) = g/[4\cdot(c - U(x))^2], \quad \Omega_0(x) = g/[4\cdot(c - U(x))].$$

**Conclusions**: The resulting form of the asymptotic of " elementary solution" points to the fact of seizure the gravity waves of certain frequencies by the flow of a given profile, while the waves of other frequencies are being oppressed, ie, one can speak of the eigen frequencies inherent to the current flow profile, and maintaining these frequencies by the flow energy at a certain time frame. Thus, the current flow acts as a frequency filter for gravity waves on the water.

28.08.2010.